\documentclass[letterpaper,twocolumn,aps,prb,showpacs]{revtex4}

\usepackage{graphicx}

\begin{document}
\title{Stacking-fault energies for Ag, Cu, and Ni from empirical
tight-binding potentials}
\author{R. Meyer}\email{ralf@thp.uni-duisburg.de}
\altaffiliation{Present address: Institut f\"{u}r Physik,
  Gerhard-Mercator-Uni\-ver\-si\-t\"{a}t Duisburg, Lotharstra{\ss}e 1,
  D-47048 Duisburg, Germany}
\author{L. J. Lewis}\email{Laurent.Lewis@UMontreal.CA}
\affiliation{D\'{e}partement de physique et Groupe de recherche en
physique et technologie des couches minces (GCM), Universit\'{e} de
Montr\'{e}al, C.\,P. 6128 Succursale Centre-ville, Montr\'{e}al
(Qu\'{e}bec) H3C 3J7, Canada}

\begin{abstract}
The intrinsic stacking-fault energies and free energies for Ag, Cu,
and Ni are derived from molecular-dynamics simulations using the
empirical tight-binding potentials of Cleri and Rosato [Phys. Rev. B
\textbf{48}, 22 (1993)]. While the results show significant deviations
from experimental data, the general trend between the elements remains
correct. This allows to use the potentials for qualitative comparisons
between metals with high and low stacking-fault energies. Moreover,
the effect of stacking faults on the local vibrational properties near
the fault is examined. It turns out that the stacking fault has the
strongest effect on modes in the center of the transverse peak and
its effect is localized in a region of approximately eight monolayers
around the defect.
\end{abstract}
\date{July 12, 2002}
\maketitle

\section{Introduction}
The stacking-fault energy is an important property of a material since
it determines to a high degree its deformation and failure
behavior.\cite{Hirth:68a} Therefore, for the computer modeling of
processes involving plastic deformation and/or failure, a correct
representation of the stacking-fault energy by the employed model is
desirable. However, many models have been constructed without
consideration of stacking-fault energies and the predicted values 
deviate often considerably from experimental results.\cite{Heino:99a,
Zimmerman:00a, Mehl:00a} A precise knowledge of the stacking-fault 
energies predicted by empirical models is therefore of high interest
in order to correctly interpret results of computer simulations.

The empirical tight-binding potentials of Cleri and Rosato
\cite{Cleri:93a} are extensively used in atomistic simulations since
they give a reasonable description of many structural and thermal
properties of fcc metals. In this work we study the intrinsic
stacking-fault energies and free energies predicted by these
potentials for Ag, Cu, and Ni and compare them with the results of
similar models for Cu and Ni. The stacking-fault energies were
calculated at $T=300$\,K from molecular-dynamics simulations and at
$T=0$\,K with the help of relaxation calculations (molecular-statics).
The derivation of the vibrational density of states (VDOS) from the
simulations made it also possible for us to calculate the
stacking-fault free energies within the quasiharmonic approximation
and to study the local effect of stacking-faults in fcc metals. 

The aim of this paper is to provide the stacking-fault energies for
Ag, Cu, and Ni modeled by the potentials of Cleri and Rosato. We
expect that these values will contribute to proper discussions of
results based on these potentials. Although in the case of Ag and Cu
the values derived from the simulations are significantly lower than
the experimental results, the trend between all three elements is well
reproduced. It has to be noted that the occurrence of low
stacking-fault energies is a general problem of central-force
potentials. One reason for this is that without the consideration of
angular forces significant contributions to the stacking-fault energy
arise from the third and higher neighbor shells,
only.\cite{Girshick:98a} A related explanation is given by the nearest
neighbor d-band tight-binding model where the energy difference between
the fcc and hcp structure is dominated by the fifth and sixth moment
of the density of states,\cite{Sutton:93a} whereas many empirical
potentials, including those of Cleri and Rosato, are based on the
second moment approximation. Better agreements of computationally
simple empirical potentials with experimental values for the
stacking-fault energy require therefore the explicit consideration of
the stacking-fault energy during the construction of the potentials.

\section{Computational Methods}
\subsection{General}
As stated in the introduction, the stacking-fault energies and free
energies reported in this article have been derived from
molecular-dynamics simulations employing the potentials of Cleri and
Rosato.\cite{Cleri:93a} In these simulations the equations of motions
have been integrated with the help of the velocity form of the
Verlet-algorithm \cite{Allen:91a} using a time-step of 2\,fs. 
Isothermic-isochoric (NVT ensemble) and isobaric-isoenthalpic
(NpH ensemble) conditions were achieved where necessary by the
application of the Nos\'{e}-Hoover thermostat \cite{Hoover:85a} method
and the Parrinello-Rahman scheme \cite{Parrinello:80a} (restricted to
orthogonal simulation boxes), respectively. Periodic boundary
conditions were applied in all cases.

Calculations of the relaxed structures at $T=\text{0\,K}$ were done in
two steps. First, the systems were cooled down to a temperature of
$T=\text{0.1\,K}$. Afterwards, the final configurations
were obtained with the help of a steepest-decent method. 

For each of the three elements a stacking-fault configuration and a
reference configuration were prepared. The stacking-fault
configurations were composed as a stack of 110 triangular lattice layers 
building a nearly cubic system of $N_\mathrm{st} = 1\,018\,160$ atoms
with one stacking-fault; the reference configurations on the other
hand are regular fcc systems$N_\mathrm{fcc}=1\,000\,188$ atoms ($63^3$
cubic cells).

At first glance, configurations with more than a million atoms may
seem to be exaggerated for the simple calculation of stacking-fault
energies. However, these system sizes guarantee the presence of a
sufficiently large number of vibrational modes for the calculations of
free energies in the quasiharmonic approximation. In addition, the
large system sizes lead to large distances between the periodic copies
of the stacking-fault, thereby minimizing possible self-interaction
effects. 

Before the calculation of the stacking-fault energies could begin, the
equilibrium dimensions of all configurations had to be determined
at $T=0$ and 300\,K. In the case of the reference configurations these
were derived from the $T=\text{0\,K}$ lattice
constants given in Ref.~\onlinecite{Cleri:93a} and the average system
volumes obtained from isobaric-isoenthalpic simulations at
$T=\text{300\,K}$ and zero pressure. 

For the determination of the equilibrium sizes of the stacking-fault 
configurations, it had to be taken into account that only volume
relaxations perpendicular to the plane of the stacking fault are
physically relevant. Relaxations in the directions parallel to the
fault-plane are artifacts of the finite system size and diminish if
the number of atomic layers parallel to the fault is increased. For
this reason the in-plane dimensions of the simulation boxes of the
stacking-fault configurations were derived from the lattice constants
obtained for the perfect systems and only the out-of-plane dimensions
had to be explicitly determined. At $T=\text{300\,K}$ these missing
dimensions were again obtained with the help of zero-pressure 
isobaric-isoenthalpic simulations in which only the out-of plane
direction of the simulation box was allowed to fluctuate. Similarly
the volume relaxations at $T=\text{0\,K}$ were derived from relaxations
of the stacking-fault configurations with fixed in-plane dimensions of
the simulation box.

\subsection{Calculation of stacking-fault energies}

After the dimensions of the configurations had been determined, the
stacking-fault energies $\gamma$ of the three metals could be
obtained. For this the simple formula
\begin{equation}
  \gamma = \frac{N_{\text{sf}} ( e_{\text{sf}} - e_{\text{fcc}} )}{A} ,
  \label{EqSF}
\end{equation}
was used, where $A$ is the area of the stacking-fault plane,
$N_{\text{sf}}$ the number of atoms in the stacking-fault
configuration and $e_{\text{sf}}$, $e_{\text{fcc}}$ the potential
energies per atom in the respective configuration. The kinetic energy
does not contribute to the stacking-fault energy since its per
particle value does not depend on the configuration.

Particular attention has to be paid to the system temperature. The
usual method to attain the desired temperature in molecular-dynamics
simulations is to rescale the particle velocities repeatedly until the
system is equilibrated. In simulations run under conditions of
constant energy or enthalpy, this procedure may lead to a small offset
in the temperature which in turn affects the value of the potential
energy. While these discrepancies usually are very small and can be
neglected in most cases, they can have an dramatic impact on 
the values of $\gamma$ since the potential energy difference on the
right hand side of Eq.~(\ref{EqSF}) is much smaller than the  
potential energies themselves and can be comparable to --- or even
smaller than --- the error in the potential energies induced by
temperature deviations. In order to avoid this problem we derived the
potential energies from simulations in the canonical (NVT) ensemble. 

\subsection{Calculation of stacking-fault free energies}

The exact calculation of free energies from molecular-dynamics
simulations is a difficult task. In this work we employ the
quasiharmonic approximation\cite{Lewis:90a} to derive the free
energies $F$ of the stacking-fault and reference configurations.
Within this approach the free energy is given by 
\begin{equation}
  F = E_0(V) + 3 N \int_0^\infty g (\omega) 
        \left [ \frac{\hbar\omega}{2} + \frac{1}{\beta} \log (1 -
        \mathrm{e}^{-\beta \hbar\omega}) \right ] d\omega,
  \label{EqQH}
\end{equation}
where  $\beta$ is the inverse temperature $(k_\mathrm{B}T)^{-1}$, 
$g(\omega)$ the normalized VDOS and $E_0(V)$ denotes the ground-state
energy of the system \emph{for the equilibrium volume at temperature
$T$}. In contrast to most molecular-dynamics simulations,
Eq.~(\ref{EqQH}) takes quantum effects fully into account. 
However, since we are mainly interested in the values of the
stacking-fault energies in a classical computer-simulation context, it
is more appropriate to calculate the free energy in the classical
limit. Taking this limit and neglecting the purely quantum zero-point
fluctuations, Eq.~(\ref{EqQH}) becomes 
\begin{equation}
  F_{\mathrm{cl}} = E_0(V) + 3 N \beta^{-1} 
                \int_0^\infty g(\omega) \log(\beta \hbar \omega) d\omega. 
\label{EqCl}
\end{equation}

Given Eq.~(\ref{EqCl}) the task of calculating the free energy of the
configurations by molecular-dynamics simulations reduces to the
determination of the vibrational density of states $g(\omega)$ and the
ground-state energy $E_0(V)$. The latter quantity was obtained by a
relaxation of the systems to $T=\text{0\,K}$ using the simulation box
dimensions of $T=\text{300\,K}$. In order to derive the vibrational
density of states $g(\omega)$, simulations in the NVE ensemble were
performed at $T=\text{300\,K}$. From the trajectories of the particles
during these simulations the velocity-autocorrelation function
$\langle \mathbf{v}(t)\mathbf{v}(0)\rangle$ was derived which is
related to the normalized VDOS by\cite{Lovesey:86a}
\begin{equation}
        g(\omega) = \int_{-\infty}^{\infty} 
               \frac{\langle\mathbf{v}(t)\mathbf{v}(0)\rangle}
                    {\langle\mathbf{v}(0)\mathbf{v}(0)\rangle}
               \mathrm{e}^{i \omega t} dt.
\end{equation}
In order to reduce statistical errors, the density of states
$g(\omega)$ was averaged over ten or more simulation runs of 5\,000
steps, each.  From the density of states obtained this way the free
energies of all six configurations could be calculated with the help
of Eq.~(\ref{EqCl}). These values were then used to calculate the
intrinsic stacking-fault free energies $\tilde{\gamma}$ of the three
elements by replacing the potential energies per particle in
Eq.~(\ref{EqSF}) by the corresponding free energies per particle. 
Since at $T=\text{0\,K}$ energy and free energy are identical these
calculations had to be done only at $T=\text{300\,K}$.

\section{Results}
\subsection{Stacking-fault Energies}
In Table~\ref{TabSfe} we present the stacking-fault energies and free
energies obtained from molecular-dynamics simulations as described in
the preceding section. From this table it can be seen that the general
agreement with the experimental results is only qualitative, similar
to comparable models. The calculated stacking-fault energies (and free
energies) for Cu and Ag are much smaller than their experimental
counterparts; the calculated stacking-fault free energy of Ag at room
temperature is even negative. For Ni the situation is not clear due to
the large variation of the experimental data. However, the trend among
the three elements is well reproduced by the model. This means that
while one has to be careful with the interpretation of quantitative
results, the potentials of Cleri and Rosato are well suited for
comparisons between materials with high, low and very low
stacking-fault energies.
\begin{table}[b]
\caption{Experimental $(e)$ and theoretical $(t)$ values of
stacking-fault energies for Ag, Cu, and Ni (in $\mathrm{mJ/m^{2}}$). 
$\gamma$ and $\tilde{\gamma}$ are the stacking-fault energies and
free energies obtained in this work; OJ and VC denote the values given
in Ref.~\onlinecite{Zimmerman:00a} for the models of Oh and Johnson
\cite{Oh:88a} and Voter and Chen.\cite{Voter:87a,Voter:95a} The
experimental results are taken from the compilations of
Gallagher\cite{Gallagher:70a} and Murr.\cite{Murr:75a}}
\label{TabSfe}
\begin{ruledtabular}
\begin{tabular}{lcrrrr}
& & $T$ & \multicolumn{1}{c}{Ag} & \multicolumn{1}{c}{Cu} &
\multicolumn{1}{c}{Ni} \\  
\hline
$\gamma$         & $(t)$ &   $0\,\mathrm{K}$ &  1 & 21 & 305 \\ 
$\gamma$         & $(t)$ & $300\,\mathrm{K}$ &  0 & 15 & 262 \\
$\tilde{\gamma}$ & $(t)$ & $300\,\mathrm{K}$ & -2 & 14 & 292 \\
OJ               & $(t)$ &   $0\,\mathrm{K}$ &    & 27 &  13 \\
VC               & $(t)$ &   $0\,\mathrm{K}$ &    & 37 &  87 \\
Gallagher        & $(e)$ &                   & 22 & 55 & 250 \\
Murr             & $(e)$ & $300\,\mathrm{K}$ & 22 & 78 & 128 \\  
\end{tabular}
\end{ruledtabular}
\end{table}

In accordance with the experimental data,\cite{Murr:75a} the results
in Table~\ref{TabSfe} show a decrease of the stacking-fault energies
and free energies with increasing temperature. In the case of the
stacking-fault energies, the decrease is caused by anharmonic effects
only, since the specific heat of a purely harmonic, classical system
does not depend on structural details. In contrast to this, the
decrease of the stacking-fault free energies in Table~\ref{TabSfe}
contains harmonic and anharmonic contributions. However, due to the
nature of the quasiharmonic approximation, the anharmonic
contributions are treated only in an approximate fashion via the
volume dependence of $E_0$ and the temperature dependence of the
VDOS. This approximation gets in the way if one wants to compare the
stacking-fault energies and free energies in order to determine the
role of entropic contributions to the stacking-fault
energy. Particularly striking is the large discrepancy between the
stacking-fault energy and free energy of Ni at room temperature. Since
our results indicate that the entropy difference favors the
stacking-fault configuration, the stacking-fault free energy should be
the lower quantity, as it is the case for Cu and Ag. This shows that,
at least in the case of Ni, there are important anharmonic
contributions to the stacking-fault free energy which are not
accounted for by the quasiharmonic approximation.

\subsection{Vibrational Density of States}
\begin{figure}[t]
\includegraphics[width=7.5cm]{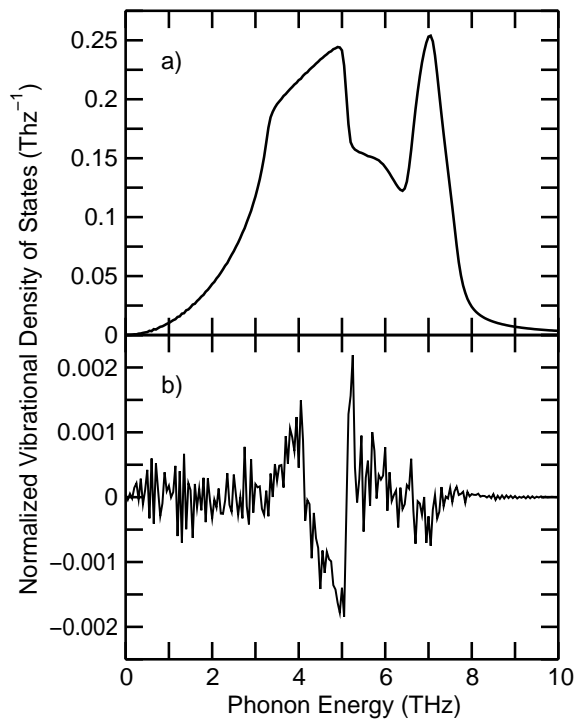}
\caption{a) Calculated normalized VDOS $g_\mathrm{fcc}(\omega)$ for
crystalline fcc Cu. b) Difference $\Delta g(\omega) = g_{\mathrm{sf}}
(\omega)-g_{\mathrm{fcc}}(\omega)$ between the VDOS $g_\mathrm{sf}
(\omega)$ of the system with stacking-fault and $g_\mathrm{fcc}
(\omega)$.}%
\label{FigDos}%
\end{figure}
Figure~\ref{FigDos} shows the VDOS calculated for the fcc structure
as well as the difference of the VDOS between the stacking-fault and
the perfect fcc configuration. It should be noted that the absolute
value of this difference is physically irrelevant since it depends
on the system size and goes to zero in the limit of an infinitely thick
system. However, the shape of this function is system size independent
and reveals the energy regions where the VDOS is affected by the
stacking fault. From the lower panel of Fig.~\ref{FigDos} it can be
seen that the strongest effect of the stacking fault appears in the
transverse peak where modes are transferred from the center of the
peak to (mostly) lower energies. A similar but smaller transfer takes
place in the longitudinal peak. It is these transfers which give rise
to the entropy difference between the configurations, which in turn
accounts for the temperature dependent change of the stacking-fault
free energy within the quasiharmonic approximation. A similar
behavior is found for Ag and Ni with a slightly more pronounced
transfer of longitudinal modes in the case of Ni.  

\begin{figure}
\includegraphics[width=7.5cm]{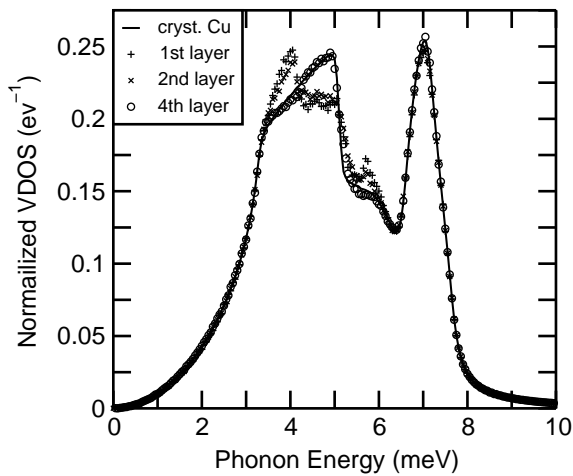}
\caption{Normalized local VDOS of the atoms in the first, second, and
fourth layer away from a stacking fault in Cu compared to the VDOS
in the perfect crystal.}%
\label{FigLayer}%
\end{figure}

In order to see the spatial extent of the influence of stacking faults
on the vibrational properties of the crystal lattice we have
calculated the local VDOS for the first atomic layers near a stacking
fault in Cu. The results of this calculation, which are presented in
Fig.~\ref{FigLayer}, show that atoms in the first and second layer near
the fault account for most of the change of the total VDOS in
Fig.~\ref{FigDos}, whereas the VDOS in the fourth layer differs only
slightly from the bulk VDOS. Atoms in the third layer (whose VDOS is
not given in order to avoid overcrowding of the figure) have an
intermediate character. While their VDOS is still clearly different
from the VDOS of the crystalline bulk, it is already much closer to
this than to the VDOS of the first two layers.  

\section{Summary and Conclusions}
The stacking-fault energies predicted by the empirical tight-binding
potentials of Cleri and Rosato reflect the experimental values of Ag,
Cu, and Ni qualitatively. Similar to other empirical models they
underestimate the values of Ag and Cu. In contrast to this and in
rough agreement with experimental data, rather high stacking-fault
energies are found for Ni. Since the experimental data for Ni show
strong variations it is impossible to say whether the potential for
this metal overestimates the stacking-fault energy or not. In any
case, the potentials can be used for qualitative comparisons of 
materials with high, low, and very low stacking-fault energies.

For all three metals we find a decrease of the stacking-fault energy
with increasing temperature of the same order of magnitude as it is
experimentally observed.\cite{Murr:75a} This decrease has to be
considered in the construction or evaluation of models which
explicitly take the stacking-fault energies into account.

The discrepancy between the calculated room temperature stacking-fault
energy and free energy for Ni shows that important anharmonic effects
are missed by the quasiharmonic approximation. For this reason 
one should not put too much emphasis on the negative value found for
the room temperature stacking-fault free energy of Ag. What remains is
the question why the anharmonic effects are so much stronger for Ni. A 
possible explanation might be given by the relative importance of the
many-body term vs. the pair-potential term in the model. An inspection
of the parameters of the potentials in Ref.~\onlinecite{Cleri:93a}
shows indeed that the pair-potential term has a lower weight and a
shorter range for Ni than for Cu or Ag. Thus, the increased weight
of the many-body term in the case of Ni might account for the increased
occurrence of anharmonic effects in this metal.

Finally, the analysis of the local effect of stacking-faults on the
vibrational properties in fcc metals reveals a strong transfer of
modes to lower energies in the transverse peak and a similar but less
pronounced transfer in the longitudinal peak of the VDOS. Inspection
of the layer-resolved local VDOS shows that these effects are
localized in a region of six to eight monolayers around the defect.

\begin{acknowledgments}
This work has been supported by grants from the Canadian
{Natural Sciences and Engineering Research Council} (NSERC),
Qu\'{e}bec's \textit{Fonds pour la formation de chercheurs et l'aide
\`{a} la recherche} (FCAR) and Germany's \textit{Deutsche
Forschungsgmeinschaft} (DFG). Most of the calculations have been
performed on the facilities of the \textit{R\'{e}seau 
qu\'{e}b\'{e}cois de calcul de haute performance} (RQCHP).
\end{acknowledgments}

\end{document}